\documentclass[]{aa}
\usepackage{graphics}

\def\cm2{cm$^{-2}$}

\begin{document}
\thesaurus{03 (11.07.1;12.03.3)}
   \title{The VLT observations of the HDF--S NICMOS field:  photometric catalog and high redshift galaxy candidates}

\titlerunning {The VLT observations of the HDF--S NICMOS field}
   \author{ A. Fontana  \inst{1} \and 
S. D'Odorico \inst{2} \and 
R. Fosbury \inst{3} \and 
E. Giallongo \inst{1} \and 
R. Hook \inst{3} \and 
F. Poli  \inst{1} \and 
A. Renzini \inst{2} \and 
P. Rosati \inst{2} \and 
R. Viezzer \inst{2} }

\institute{  
  Osservatorio Astronomico di Roma, via dell'Osservatorio, I-00040
Monteporzio, Italy 
\and
  European Southern Observatory, Karl Schwarzschild Strasse 2,
D-85748 Garching, Germany
\and
Space Telescope - European Coordinating Facility, Karl Schwarzschild Strasse
2, D-85748, Garching bei Muenchen, Germany.}

   \offprints{A. F.: {\sf fontana@quasar.mporzio.astro.it}}
   \date{Received:December 1, 1998; Accepted: December 11, 1998 }

   \maketitle

   \begin{abstract} We present the deep $UBVRI$ observations of the HDF-S NICMOS
	field obtained as part of the
	Science Verification of the VLT Unit 1 telescope. The images have been
	used to construct  object catalogs and to obtain photometric
        redshifts. The effective field of view is $\simeq
	70\times70$ arcsec$^2$, and the
	formal $5\sigma$ limiting magnitudes (in a 2 FWHM aperture)
	are 26.3, 27.8, 27.5, 26.9, 25.2 in the $U$, $B$, $V$, $R$
	and $I$ bands,
	respectively.  Thanks to the sub-arcsecond image quality,
	relatively long exposure time, and large collecting area of the
	VLT, this is the deepest set of multicolor images ever
	obtained from a ground--based telescope.

	Galaxy counts have been derived independently in each band,
	and show no significant departures from previous data from
	wider areas. A multicolor photometric catalog of all the
	galaxies selected in the $R$ band has also been obtained and
	used to derive photometric redshifts for all galaxies with
	$R \leq 26.5 $, using also the J,H and K magnitudes from the
	NICMOS deep observations.  A significant fraction
	($\simeq$28\%)of the galaxies is placed at $z\geq 2$.  Among
	them, six robust candidates are found at $z \geq 2.5$ .
   \end{abstract}
\section{Introduction}

Deep imaging of extragalactic fields has long been recognized to be a
powerful tool to understand galaxy evolution (see Ellis 1997 for an
extensive review). Although faint galaxy counts have been of paramount
importance to show that galaxies do evolve with redshift, the overall
scenario and the physical processes that led to galaxy evolution are
still debated.  A new approach developed in recent years uses deep
{\it multicolor} surveys to study the fainter magnitude galaxies: deep
multi-band images are taken with a complete set of broad--band filters,
in order to cover the overall spectrum of the galaxy and to
discriminate the populations at different redshifts.  The Hubble Deep
Field North (Williams et al. 1996) is the best--known examples of this kind
of observations, but  ground--based images have also been used, mainly to
define sharp color criteria that select high--redshift galaxy
candidates (Steidel et al 1995, Giallongo  et al. 1998, Arnouts et al
1998).

This paper analyses deep observations of the Hubble Deep
Field-South (HDF-S, Williams et al 1999) obtained in five colors
($U$, $B$, $V$, $R$ and $I$) in August 1998 as part of the Science 
Verification phase of the first VLT 8.2m
telescope (UT1). A description of  the Science Verification programme
at the VLT is to be found at\\
{\sf http://www.eso.org/paranal/sv/} .

The data have been taken for a field
centered on $\alpha= 22h 32m 51.7$, $\delta= -60d 38` 48.2''$ (J2000),
thus providing the optical complement to the near-IR $J$, $H$ and $K$
band images obtained with the HST NICMOS instrument (Fruchter et al 1999,
see also \\
{\sf http://www.stsci.edu/ftp/science/hdfsouth/hdfs.html}), since
the area covered by the optical observations is somewhat more extended
than the NICMOS field of view.

This paper is organized as follows: In Section 2 we describe the data
reduction procedures, while in Section 3 we discuss how the
photometric catalogs were constructed. In Section 4 we derive the
photometric redshifts for all the objects brighter than $R\simeq 26.5$
and discuss briefly their distribution at high redshift.

\section{The Data Sample}
\subsection{Observations and data reduction}

 \begin{table*}	
\caption{Summary of the Observational Data} \label{tobs}
\begin{tabular}{cccccccc}
\hline
Filter &  Number of  &  total & Final PSF (FWHM) & Field of View & Zero-point
& Limiting Mag. \\
       & frames & exp. time (s) & (arcsec)  & (arcsec)$\times$(arcsec) $^{(1)}$ & $^{(1)}$ & (5 $\sigma$) $^{(2)}$ \\
\hline

$U$     & 16 & 17800 & 0.73 & 68.8 $\times$ 72.8 & 31.58 &  26.3 \\
$B$     & 15 & 10200  & 0.78 & 71.7$\times$71.7 & 33.13 & 27.8 \\
$V$     & 16 & 14400  & 0.82 & 78$\times$69.5 & 33.85 & 27.5 \\
$R$     & 8 & 7200  & 0.64 & 72.6$\times$ 66.9 & 34.06 & 26.9 \\
$I$     & 12 & 10160 & 0.71 & 75.6$\times$ 67.0 & 33.44 & 25.2 \\
\hline
\end{tabular}

(1) : Referred to the images released by ESO\\ 
(2) : Magnitude limits are computed from the standard deviation of the
sky counts inside apertures with diameters of 2$\times$ FWHM without
aperture correction.

\end{table*}
The data used here were retrieved from the ESO public archive, and
refer to the observations that were obtained in the period August
17-September 1, 1998, using the VLT Test Camera (VLTTC) at the
Cassegrain focus of the UT1. The VLTTC is a simple imaging camera 
which reimages the focal plane onto a $2048^{2}$, 24 $\mu$m pixels,
thinned SITe CCD. In the $2\times 2$ binned mode which has been used
for these observations the scale is 0.092 arcsec/pixel, giving a total
field of 92$\times$92 arcsec$^2$.
For this program a set of several exposures through the standard $UBVRI$
Johnson-Cousin filters was obtained with single exposure integration
times ranging from 600 to 1200 seconds.  Airmass of individual
exposures was always $\leq 1.4$, with a median value of 1.25.
Observations were obtained following
the standard criteria for deep imaging, i.e. applying a slight offset
between individual pointings to allow the removal of the detector
imprints.  Full details on the Test Camera,
the CCD detector, the filter curves and on the individual exposures are available on
the WEB site \\
{\sf http://www.hq.eso.org/paranal/sv}.

The data reduction has been carried out with different softwares
at ESO and at the Rome Observatory. Although the two pipelines used
completely independent environments and tools, they are quite similar
in concept. Most of the steps are identical to those described
elsewhere (e.g. Giallongo et al. 1998; Arnouts et al. 1998), and will
not be repeated here. It is worth noting that flat-fielding is
particularly critical in the VLTTC, since its CCD suffers from a very
large blemish near the center and other lesser defects in the whole
area. The central blemish is wavelength dependent, while variations
from night to night caused by moving dust grains are also noticeable
in the flats.  Nevertheless, a satisfactory solution was finally found
by constructing a separate ``super-flat'' as the median image of the
unregistered images for each observing night. The final accuracy in
flat--fielding is estimated to be better than 1\%.

Only frames with seeing better than 1 arcsec were used in the final
coaddition, without applying any drizzling algorithm (Fruchter and
Hook 1998) since the VLTTC sampling is always much smaller than the
seeing.  As is customary in dithered multiple exposures, the edges of
the final images are of poorer quality, since only a limited number of
frames contribute to the observed flux. In the present observations
the problem is noticeable because a large dithering pattern had to be
applied to remove the blemishes and because of the small size of the
field of view. Moreover, the central pointing of the coadded frames
was not exactly the same in the different bands, which has further
reduced the area common to all frames.  We therefore used two
different sets of images. In the first - that was part of the ESO
public release -- we trimmed the outer regions independently, keeping
only the inner regions that were covered with 100\% of the
observations. These images were used to extract independent catalogs
in each band.

We also produced a set of coadded images that are trimmed and aligned to the
central field of the $R$ frame, and these images were used to prepare the
multicolor catalog. In practice, about 20\% of this field
is not covered by all the $U$ and $B$ frames, and therefore the
coadded
images are slightly shallower in these bands. The recently available images
obtained by NICMOS with the F110W, F160W and F222M  filters (Fruchter et al 1999) were also rebinned and aligned  to the VLTTC $R$ frame.

Table~\ref{tobs} summarizes the observational data that have been
 used, the FWHM of the PSF of the coadded single color
images, their area, the photometric zero point, 
 together with the
formal $5\sigma$ limiting magnitudes. These were conservatively
computed by taking the
$\sigma$--clipped standard deviation of the sky counts
in an aperture 2$\times$FWHM wide, taken at random positions on the
images.

\begin{figure*}
\resizebox{12cm}{!}{\includegraphics{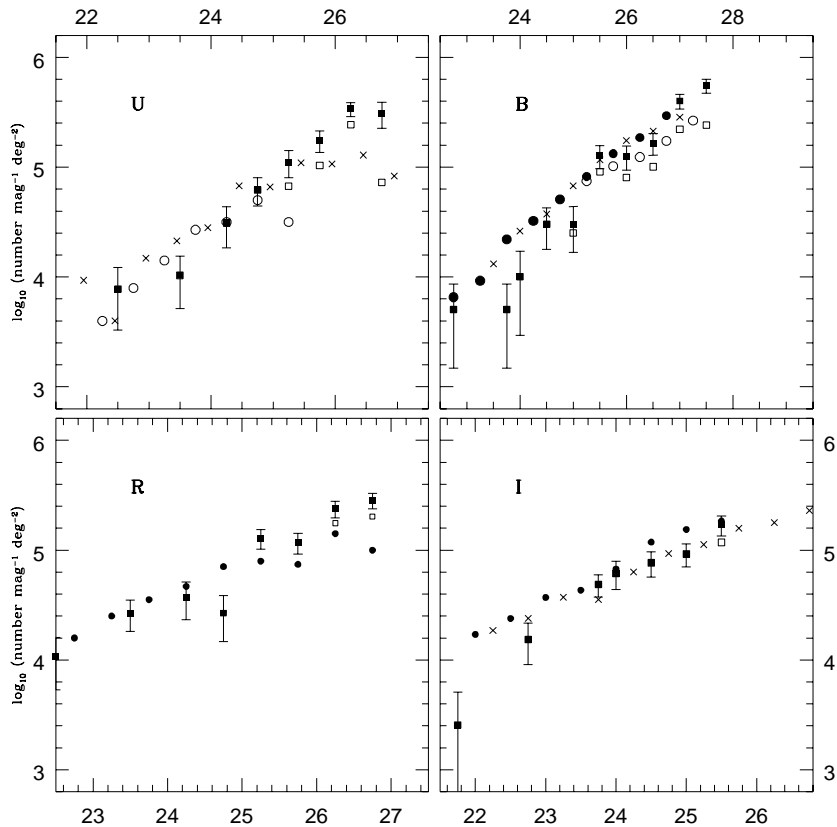}}
\hfill
\parbox[b]{55mm}{
\caption{Differential galaxy number counts for the U, B, R and I
bands.  VLT data are shown with open (raw) and filled
(corrected) squares.  Error bars take into account Poisson error only.
Other data taken from the literature are shown.
{\it U Band}: empty circles from Hogg et al 1995; crosses from HDF--N
(Williams et al 1996). A correction $U_J = U_{AB} -0.8$ has been
applied to the latter data.  {\it B Band}: (open) filled circles are
(un--)corrected data from Metcalfe et al 1995a. Crosses from Arnouts et
al 1998. {\it R Band}: Filled circles are raw Keck data from  Smail et al 1995. {\it I Band}: Filled circles are from the NTT Deep Field (Arnouts et al 1998), crosses from HDF--N (Williams et al 1996).}

\label{counts}
}
\end{figure*}

\subsection{Photometric calibration}
The photometric calibrations were obtained by reducing a series of
standard stars from the Landolt (1992) sample.  Self-consistent
photometric solutions to the standard Johnson-Cousin system were derived for
several individual nights, along with average solutions that use all
the data from photometric nights. The zero points, color terms and
extinction coefficients derived are listed at \\
{\sf
http://www.hq.eso.org/paranal/sv/html/data/photom.txt}.  We emphasize
however that the coadded
images presented here cannot be calibrated directly using
the average coefficients, since they are the average of different
exposures taken under various conditions.

A few isolated relatively bright stars have been selected in the
field. Then, images obtained during each nights for which a
photometric solution exists have been taken, and accurate 
magnitudes of the selected stars have been measured in each of these
images. The resulting instrumental magnitudes have been converted into
Johnson magnitudes using the photometric solution for the given
night. Finally, each star has been assigned  its average
magnitude, after a $\sigma$--clipping removal of the discordant
values.
 
The same  photometry  has been applied to the summed
images, and the final zero point has been chosen in order to reproduce
the magnitudes of the selected stars. 
We estimate that the final accuracy of this procedure is 
$\sim 0.05$ mags in each band.
Finally, the zero points were corrected for galactic absorption with
$E(B-V) = 0.02$ (Burstein \& Heiles 1982) with $\delta U=0.095$,
$\delta B=0.084$, $\delta V=0.063$, $\delta R=0.05$, $\delta I=0.038$.

\section {The Photometric Catalog}

The analysis of the two sets of images (the individual coadded images
and
those trimmed to the $R$ band image) was performed
using the SExtractor image analysis package (Bertin \& Arnouts, 1996)
.

\subsection{Galaxy Counts}
Within SExtractor, images were smoothed with a gaussian filter
matching the seeing, and the detection threshold was chosen at
$3\sigma$ of the background intensity in a contiguous area of 1
FWHM.   Following Djorgovski et al.
(1995), for each object both isophotal and aperture magnitudes (in a
2 FWHM aperture) were computed.  The isophotal magnitude was used for
the larger/brighter objects, i.e. for those objects where the isophotal area is
larger than the aperture one. For fainter objects, an aperture
correction to $5''$ has been estimated on bright stars and applied 
to correct the 2
FWHM aperture magnitude. This procedure is strictly valid for
star-like objects only, but has been shown to be a good approximation
on deep images (Smail et al. 1995).

Bright stars have been excluded from the catalogs using the
CLASS\_STAR parameter provided by SExtractor. A threshold
CLASS\_STAR$ < 0.9$ has been set, on the basis of the comparison between 
ground--based and HST data (Arnouts et al 1998). 
At fainter magnitudes (e.g. $R\geq 24$)
the neural network classifier does not work properly anymore, but
stars are not expected to dominate the counts and therefore
have been ignored. All the single-color catalogs are available on the web
site \\
{\sf
http://www.mporzio.astro.it/HIGHZ}.

A correction for incompleteness has been estimated in each band as in
Arnouts et al. (1998), accounting for both false detection and the
non--detection of real objects. We warn the reader that the correction
for incompleteness has to be taken with some caution, because of the
very limited size of the sample used as a reference.
The raw and corrected counts in each band are
shown in Fig~\ref{counts}, and compared to the most recent data from
the literature.  It is noteworthy here that the counts derived from
these data are the deepest ever obtained from a single ground-based telescope,
thanks to the sub-arcsecond image quality, relatively long exposure
time and the large collecting area of the VLT.

\subsection{The Multicolor Catalog}

The multicolor catalog has been obtained from the set of aligned
images, taking the $R$ frame as reference. Object detection and the
measurement of the $R$ magnitude have been performed on the $R$ frame
exactly as described in the previous section. Then colors have been
measured in a fixed circular aperture of 14 pixels (corresponding to 2
FWHM of the $R$ frame), keeping the object position found on the $R$
frame. To allow for the  seeing difference among the coadded
images, colors have been measured on images degraded to the $0.82''$
seeing of the $V$ band image. The $R\leq 26.5$ subsample used for the
following analysis consists of 91 galaxies and is available on the WEB
\\{\sf http://www.mporzio.astro.it/HIGHZ}.  The $R=26.5$ threshold has
been set in order to ensure good photometric accuracy and meaningful
colors on the whole sample. Obvious stars have been excluded using the
CLASS\_STAR parameter (see above) in the R and - when available - 
in the NICMOS images.  

\begin{figure}
\resizebox{\hsize}{!}{\includegraphics{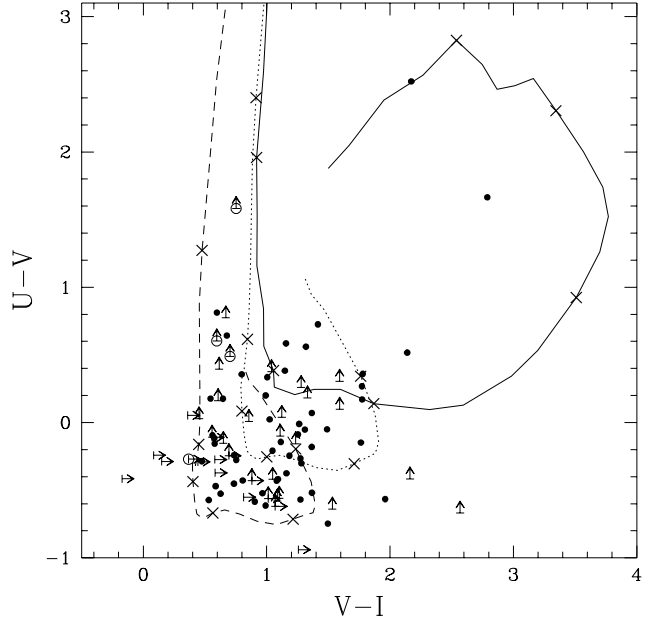}}
\caption{The color distribution of galaxies in the VLT field.  Points
refer to galaxies in the $R\leq26.5$ sample. Lower limits are shown at
the 3$\sigma$ level. Circles marks the objects at $z\geq 2.5$ selected
from the photometric redshift.  Overplotted are the evolutionary
tracks of three galaxy templates, computed with a Miller--Scalo IMF
and the following parameters: $\tau=0.3$ Gyr, $z_{form}=7$ and
E(B-V)=0.1 (solid line); $\tau = 5$ Gyr, $z_{form}=5$ and
$E(B-V)=0.2$(dotted line); $\tau = \infty $, $z_{form}=5$, no dust (dashed
line). Points at $z=0.5,1,1.5,2,2.5,3$ are marked by a cross.  }
\label{colors}
\end{figure}

\begin{figure}
\resizebox{\hsize}{!}{\includegraphics{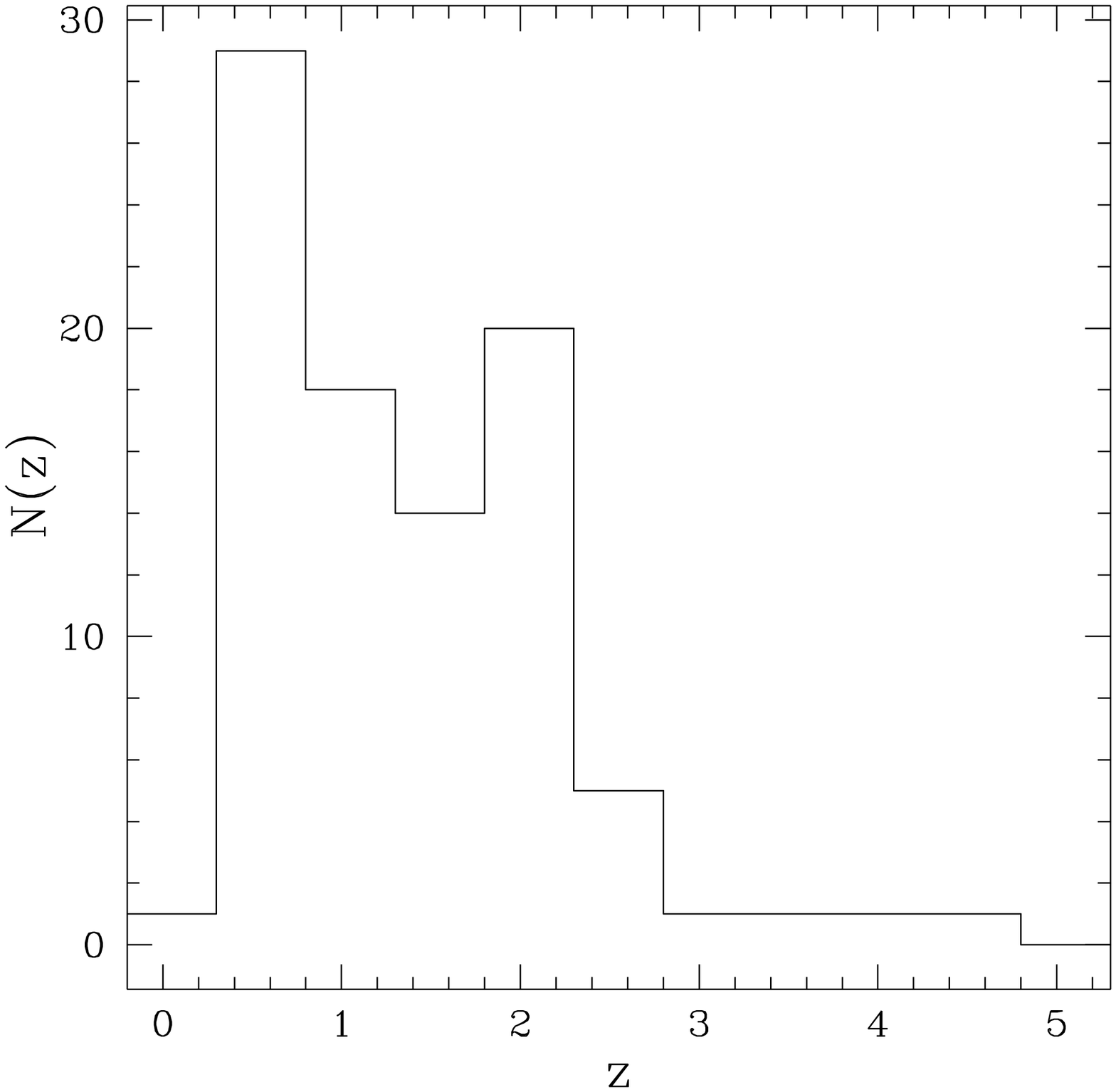}}
\caption{The photometric redshift distribution of galaxies at $R\leq
26.5$ in the VLT field.}
\label{nz}
\end{figure}

Fig~\ref{colors} shows the color distribution of the $R\leq26.5$
sample in the $V-I$ vs $U-V$ plane. Overplotted on the observed colors are the evolutionary
tracks of few galaxy templates as a function of redshift. They have
been  chosen to
broadly encompass the most common spectral types, and are based on
the synthetic models of Bruzual and Charlot (GISSEL library, 1996). 
It is clearly
seen that a significant fraction of the faint galaxies has colors
typical of star-forming galaxies at $z\simeq 2$, as expected in these
very deep images (Metcalfe et al. 1995b). Though most of these objects
are probably too faint for a spectroscopic confirmation with FORS or
ISAAC at the VLT, their nature and redshifts can be investigated
further by means of a photometric redshift analysis.

\section{The Photometric Redshifts of the $R\leq26.5$ galaxies}

\begin{table}	
\caption{Galaxy candidates at $z\geq 2.5$
} \label{tabhighz}
\begin{tabular}{rrrrr} 
\hline
OBJ\#   & X $^{(1)}$      & Y$^{(1)}$      & $R$ & $z_{\rm phot}$ \\
\hline
VLT  73 &  629.4 & 463.9 & 26.13 & 2.5 \\
VLT 116 &  478.1 & 237.6 & 25.32 & 2.5\\
VLT 98 & 244.4 & 332.0 & 25.44 & 3.05\\
VLT 27 & 747.6 & 619.7 & 24.29 & 3.45\\
VLT 96 & 722.2 & 339.4 & 25.07 & 3.9$^{*}$ \\
VLT 24 & 64.7 & 630.5 & 25.64 & 4.65$^{*}$ \\

\hline
\end{tabular}

(1) : Referred to the R image released by ESO\\
(*) : unresolved in the VLT images.
\end{table}

\begin{figure*}
\resizebox{12cm}{!}{\includegraphics{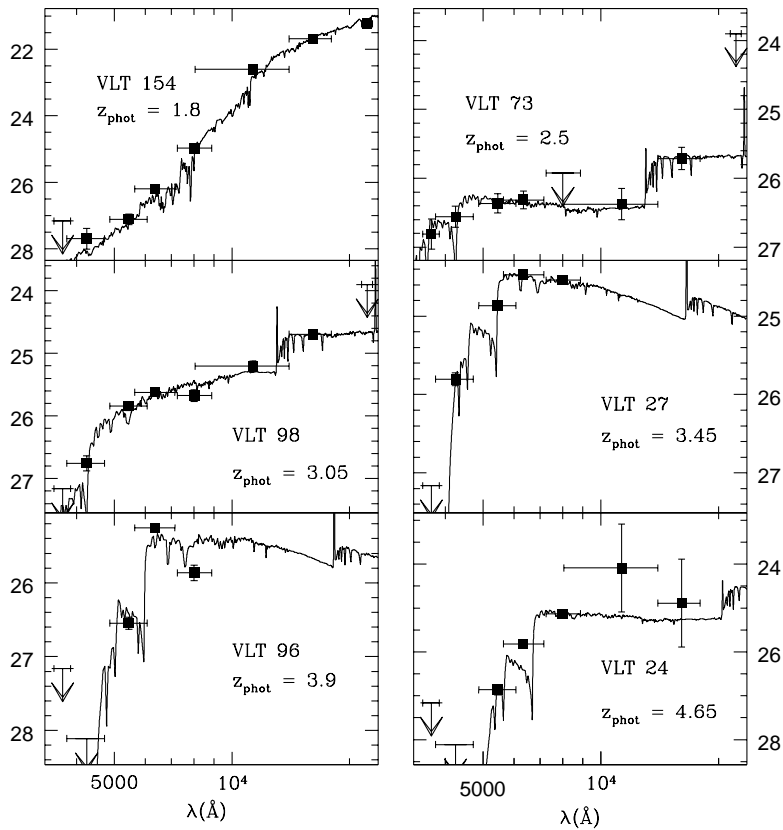}}
\hfill
\parbox[b]{55mm}{
\caption{The broad band energy distributions of high redshift galaxy
candidates. Object VLT 154 was found by Treu et al 1998. Other
objects are galaxies at $z\geq3$
in the VLT field. Magnitudes have been converted into the AB system by
$U_{AB}=U + .71$, $B_{AB}=B - 0.09$, $V_{AB}=V- 0.003$, $R_{AB}=R +
.18$, $I_{AB}=I + .42$.  The best--fitting spectrum is also shown.}
}
\label{highz}

\end{figure*}

The multicolor catalog has been used to derive  photometric redshifts for all
the 91 galaxies brighter than $R=26.5$, using a technique extensively
described elsewhere (Giallongo et al. 1998). In brief, we have
computed the expected galaxy colors 
as a function of redshift for synthetic models in the 
GISSEL library, for an extensive variety of combinations of age,
metallicity,
IMF, and e-folding star formation time-scale. 
The reddening produced by internal dust in star-forming galaxies has
then been added using the SMC extinction law (Pei 1992), along with the  absorption 
produced by hydrogen in the intergalactic medium (Madau 1995).
Finally, at any redshift  galaxies are allowed to have any age smaller
than the Hubble time at that redshift ($\Omega =1$ and $H_0 = 50$ km
s$^{-1}$Mpc$^{-1}$ have been adopted throughout the paper).

The resulting large
dataset includes $\simeq 5\times 10^5$ ``simulated galaxies'',
 and a classical $\chi ^2$--minimization procedure has been applied
to find the best-fitting spectral template to the observed colors.

This procedure has been tested on 108 galaxies with spectroscopically confirmed
redshifts in the HDF-N (Cowie 1997; Cohen et al. 1997; Dickinson et
al. 1997; Lowenthal et al. 1997 Fernadez--Soto et al 1998), obtaining
an accuracy $\sigma _z\sim 0.1$ in the redshift interval $z=0-3.5$
(Giallongo et al. 1998; Fontana et al. 1999).

The resulting redshift distribution is shown in Fig~\ref{nz}. A peak
is clearly seen at $z=0.5-1$, with a well populated tail extending to
higher redshifts: about 28\% of galaxies result indeed at $z\geq 2$.

A very red object was already identified as a $z\simeq2$ candidate
from the preliminary NICMOS observations, from being detected in the
$H$ band (F160W) and undetected in CTIO 4m telescope $R$ and $I$ band
images with $(R-H)_{\rm AB}>3.9$ and $(I-H)_{\rm AB}>3.5$ (Treu et al
1998).  This object (named VLT~154 in our catalog) 
is detected in all the VLT images except the $U$
band, and our photometric redshift analysis indicates a redshift
$z\simeq 1.8$ (see Fig~\ref{highz}).  The redshift accuracy is limited
by the lack of major spectral features, and acceptable solutions in
the range $1.5 < z < 2.1$ can be found using different combinations of
the parameters involved, consistent with the result of Stiavelli et
al. (1998).  The $z=1.8$ best-fit is obtained with solar metallicity,
$E(B-V)=0.15$ (with a SMC extinction law), star--formation
timescale $\tau=0.3$ Gyr and an age of 2~Gyrs.  Adopting the Calzetti (1997)
extinction law we obtain $z=1.65$ with $E(B-V)=0.2$. Assuming no
reddening from dust, we obtain an even higher best--fit redshift of
$z=2.05$. At $z=1.8$ , this object would have $M_K=-25.67$ and
$M_B=-21.84$ ($k$--corrections are computed exactly from the
best--fitting spectrum). Spectroscopic follow--up with ISAAC will
hopefully reveal whether this object is indeed a high redshift
elliptical galaxy undergoing
passive evolution, as suggested by the fit parameters, 
a result that has a wide cosmological relevance.

The list of the objects at $z\geq2.5$ is given in
Table~\ref{tabhighz}, while Fig.~\ref{highz} shows the best fitting
spectra of five  of them. Two of these objects are unresolved in the\
VLT images, although they are too faint for the CLASS\_STAR parameter
to be reliable. Since they also fall outside 
the HST- STIS image overlapping the HDF-S NICMOS field, we are not
able to exclude the possibility that they are actually
galactic stars, that are the major source of
interlopers in the Steidel et al. (1996) sample.

Before comparing these results to the HDF-N it is worth noting that
HDF F300W filter is significantly wider and bluer than the Johnson $U$
used here, with only a small overlapping region. As a result, the
redshift range sampled by the HDF is wider and centered at a lower
redshift ($2 < z < 3.2$) than in the present work ($2.5 < z < 3.4$),
and the surface density of ``$U$-dropout'' galaxies in HDF-N with
$V_{606}\leq 26.5$ is significantly higher than found here, or $\simeq
18$ arcmin $^{-2}$ (Pozzetti et al. 1998).

Only one galaxy candidate at $z\geq4$ results from the
redshift distribution in Fig.~\ref{nz}. This is consistent with the
number density of 0.8 arcmin$^{-2}$ ``$B$-dropout" galaxies detected in
the HDF-N (Pozzetti et al. 1998) and lower than a similar prediction in
the NTT-Deep Field (2.7 arcmin$^{-2}$ at $r\leq$26, Arnouts et al 1998).  

Given the small size of the field studied with the VLT-TC
these results are obviously  of limited
statistical significance. They demonstrate however the potential of 
 future, wide field VLT instruments like FORS-1 and FORS-2, now 
planned to become operational in 1999 and 2000. These instruments
will allow to explore to an unprecedented depth 
the  distribution and evolutionary status of galaxies in the early universe.

\begin{acknowledgements}
We thank B. Leibundgut for obtaining most of the observations,
W.Freudling for providing a preliminary NICMOS image of the field and 
G. De Marchi,  F. Natali and V. Testa for their
help in the data analysis .
R.F. is affiliated to the Astrophysics Division, Space Science Department,
European Space Agency.

\end{acknowledgements}


\begin{thebibliography}{}
\bibitem{}{Arnouts, S., D'Odorico, S., Cristiani, S., Zaggia, S., Fontana, A., 
Giallongo, E. 1998, A\&A}
\bibitem{}{Bertin,\, E., Arnouts, S. 1996, A\&AS, 117, 393}
\bibitem{}{Burstein, D., Heiles, C., 1982, AJ, 87, 1165}
\bibitem{}{Calzetti, D. 1997, in ``The Ultraviolet Universe at Low and High
	Redshift'', astro-ph/9706121}
\bibitem{}{Cohen et al  1996, ApJ, 471, L5}
\bibitem{}{Cowie, L. L. 1997, http://www.ifa.hawaii.edu/cowie/tts/tts.html}
\bibitem{}{Dickinson 1997, in the Hubble Deep Field, proceedings of the Space telescope Science Institute 1997 May Sympozium, eds. M. Livio, S.M. Falls, and P. Madau}
\bibitem{}{Djorgovski S.G., Soifer, B., Pahre, M. et al., 1995, ApJ, 438, L13} 
\bibitem{}{Ellis, R. S., 1997, ARAA, 389}
\bibitem{}{Fernandez--Soto, A., lanzetta, K., Yahil, A., 1998, astro-ph 9809126}
\bibitem{}{Fontana, A. et al 1999, in prep.}
\bibitem{}{Fruchter, A.S. \& Hook, R.N. 1998, in "Applications of Digital Image Processing XX" 
ed. A. Tescher, Proc. S.P.I.E vol 3164, p120}
\bibitem{}{Fruchter, A.S. et al, 1999, AJ to be submitt.}
\bibitem{}{Giallongo,E., D'Odorico, S., Fontana, A. et al., 1998, AJ, 115, 2169}
\bibitem{}{Hogg, D., Pahre, M., Mc Karthy, J., et al., 1997, MNRAS, 288, 404} 
\bibitem{} {Landolt, A.U., 1992, AJ, 104, 340}
\bibitem{}{Lowenthal, J. D., Koo, D. C., Guzman, R., Gallego, J.,
	Phillips, A. C., Faber, S. M., Vogt, N. P., Illingworth, G. D.,
	Gronwall, C. 1997, ApJ 481, 673}
\bibitem{}{Madau, P. 1995, ApJ, 441, 18 }
\bibitem{}{Metcalfe, N., Shanks, T., Fong, R., Roche, N., 1995a, MNRAS, 273, 257}
\bibitem{}{Metcalfe, N., Shanks, Campos, A., R. Fong, J.P.Gardner 1995b Nature 383, 236  XXX}
\bibitem{}{Pei, Y.C., 1992, ApJ, 395,130} 
\bibitem{}{Pozzetti, L., Madau, P., Zamorani, G., Ferguson, H.C., Bruzual, G., 1998,
MNRAS 1133}
\bibitem{}{Smail, I., Hogg, D. W., Yan, L., Cohen, J. G. 1995, ApJ, 449, L105}
\bibitem{}{Steidel,\, C. C., Pettini, M., Hamilton,\, D.,
	1995, AJ, 110, 2519}
\bibitem{}{Steidel,\, C. C., Giavalisco,\, M., Pettini,\, M.,
	Dickinson,\, M., Adelberger,\, K. L. 1996, ApJ, 462, L17}
\bibitem{}{Stiavelli, et al, A\&A submitted}
\bibitem{}{Treu et al, 1998, A\&A Letter, 340, 10}
\bibitem{}{Williams, R. E. et al. 1996, AJ, 112, 1335}
\bibitem{}{Williams, R. E. et al. 1999, AJ to be submitted}
\end{thebibliography}
\end{document}